\documentclass[sigconf,nonacm]{acmart}
\usepackage{natbib}
\usepackage{siunitx}
\usepackage{multirow}
\usepackage{makecell}
\usepackage{amsmath}
\usepackage{subcaption}
\usepackage[linesnumbered,ruled]{algorithm2e}
\usepackage[capitalise]{cleveref}

\newcommand{\bert}{\textsc{BERT}}
\newcommand{\dpr}{\textsc{DPR}}
\newcommand{\ance}{\textsc{ANCE}}
\newcommand{\colbert}{\textsc{ColBERT}}
\newcommand{\tct}{\textsc{TCT-ColBERT}}
\newcommand{\tildeone}{\textsc{TILDE}}
\newcommand{\tildetwo}{\textsc{TILDEv2}}
\newcommand{\staradore}{\textsc{STAR+ADORE}}
\newcommand{\condenser}{\textsc{Condenser}}
\newcommand{\cocondenser}{\textsc{coCondenser}}

\newcommand{\roberta}{\textsc{RoBERTa}}
\newcommand{\mpnet}{\textsc{MPNET}}
\newcommand{\clear}{\textsc{CLEAR}}
\newcommand{\coil}{\textsc{COIL}}

\newcommand{\approach}{\textsc{DAFT}}

\newcommand{\msmpsgdev}{\textsc{MSM-Psg-Dev}}
\newcommand{\trecdlpsg}{\textsc{TREC-DL-Psg}}
\newcommand{\msmdocdev}{\textsc{MSM-Doc-Dev}}
\newcommand{\trecdldoc}{\textsc{TREC-DL-Doc}}

\newcommand{\collapse}{\includegraphics[scale=0.1]{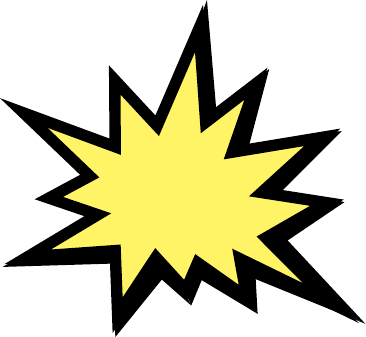}}

\DeclareMathOperator*{\kargmax}{\textit{k}\text{-}argmax}

\newcommand{\impr}[1]{\tiny ($\textcolor{green}{\blacktriangledown}#1\%)$}
\newcommand{\sig}[1]{\tiny \textsuperscript{\texttt{[#1]}}}

\setcopyright{acmcopyright}
\copyrightyear{2018}
\acmYear{2018}
\acmDOI{XXXXXXX.XXXXXXX}

\acmConference[Conference acronym 'XX]{Make sure to enter the correct
  conference title from your rights confirmation email}{June 03--05,
  2018}{Woodstock, NY}

\acmPrice{15.00}
\acmISBN{978-1-4503-XXXX-X/18/06}

\begin{document}
\title{Distribution-Aligned Fine-Tuning for Efficient Neural Retrieval}

\author{Jurek Leonhardt}
\authornote{Both authors have contributed equally.}
\affiliation{
  \institution{L3S Research Center}
  \city{Hannover}
  \country{Germany}
}
\email{leonhardt@L3S.de}

\author{Marcel Jahnke}
\authornotemark[1]
\affiliation{
  \institution{L3S Research Center}
  \city{Hannover}
  \country{Germany}
}
\email{jahnke@L3S.de}

\author{Avishek Anand}
\affiliation{
  \institution{Delft University of Technology}
  \city{Delft}
  \country{Netherlands}
}
\email{avishek.anand@tudelft.nl}

\begin{abstract}
Dual-encoder-based neural retrieval models achieve appreciable performance and complement traditional lexical retrievers well due to their semantic matching capabilities, which makes them a common choice for hybrid IR systems. However, these models exhibit a performance bottleneck in the online query encoding step, as the corresponding query encoders are usually large and complex Transformer models.

In this paper we investigate \emph{heterogeneous} dual-encoder models, where the two encoders are separate models that do not share parameters or initializations. We empirically show that heterogeneous dual-encoders are susceptible to \emph{collapsing representations}, causing them to output constant trivial representations when they are fine-tuned using a standard contrastive loss due to a distribution mismatch. We propose \approach{}, a simple two-stage fine-tuning approach that aligns the two encoders in order to prevent them from collapsing. We further demonstrate how \approach{} can be used to train efficient heterogeneous dual-encoder models using lightweight query encoders.
\end{abstract}




\maketitle
\section{Introduction}
\label{sec:intro}
Neural language models are a popular choice for a large variety of IR tasks nowadays. A common architecture comprises a \emph{query encoder} and a \emph{document encoder}, both of which map their respective input strings to fixed-size vector representations. The relevance of a query-document pair is then computed as the dot product of the query and document representation vectors. This is referred to as \emph{two-tower}, \emph{bi-encoder} or \emph{dual-encoder} architecture and has been used for retrieval~\cite{karpukhin2020dense,lindgren2021efficient,johnson2021billion} and re-ranking~\cite{zhuang2021tilde,leonhardt2022efficient,jung2022semi}. Typically, the query and document encoder either (1) are architecturally identical and initialized using the same pre-trained model or (2) even share their weights in a \textit{Siamese} fashion. We refer to such an architecture as \emph{homogeneous dual-encoders}.

Dual-encoder models complement term-matching-based retrievers well due to their semantic matching capabilities~\cite{wang2021bert}. However, as query and document encoders are usually self-attention-based Transformer models, they exhibit quadratic time complexity with respect to the input length. For the document encoder, this is usually not problematic, as the document representations are pre-computed during the indexing stage. However, the quadratic time complexity, along with the large number of Transformer layers, results in increased query processing latency primarily due to the forward pass. This means that the query encoding operation is a bottleneck, as the forward pass has to be performed during the retrieval (or re-ranking) stage, since the queries are unknown in advance.

\begin{figure}
    \begin{subfigure}{.475\linewidth}
      \includegraphics[width=\textwidth]{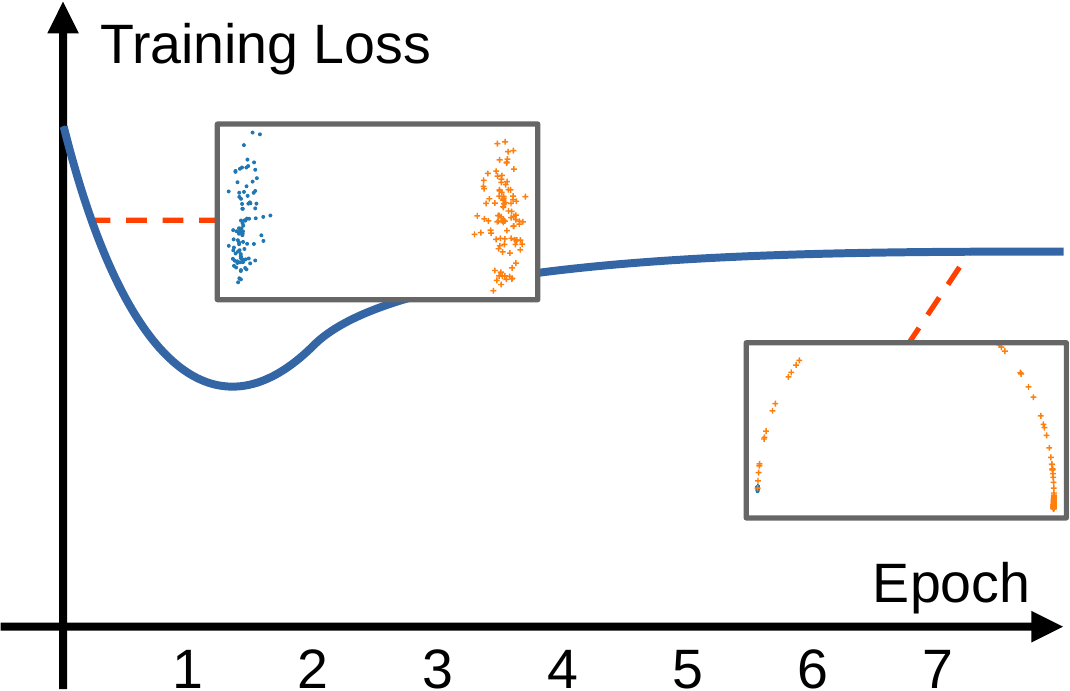}
      \caption{without alignment}
      \label{fig:intro.training_loss.no_daft}
    \end{subfigure}
    \hfill
    \begin{subfigure}{.475\linewidth}
      \includegraphics[width=\textwidth]{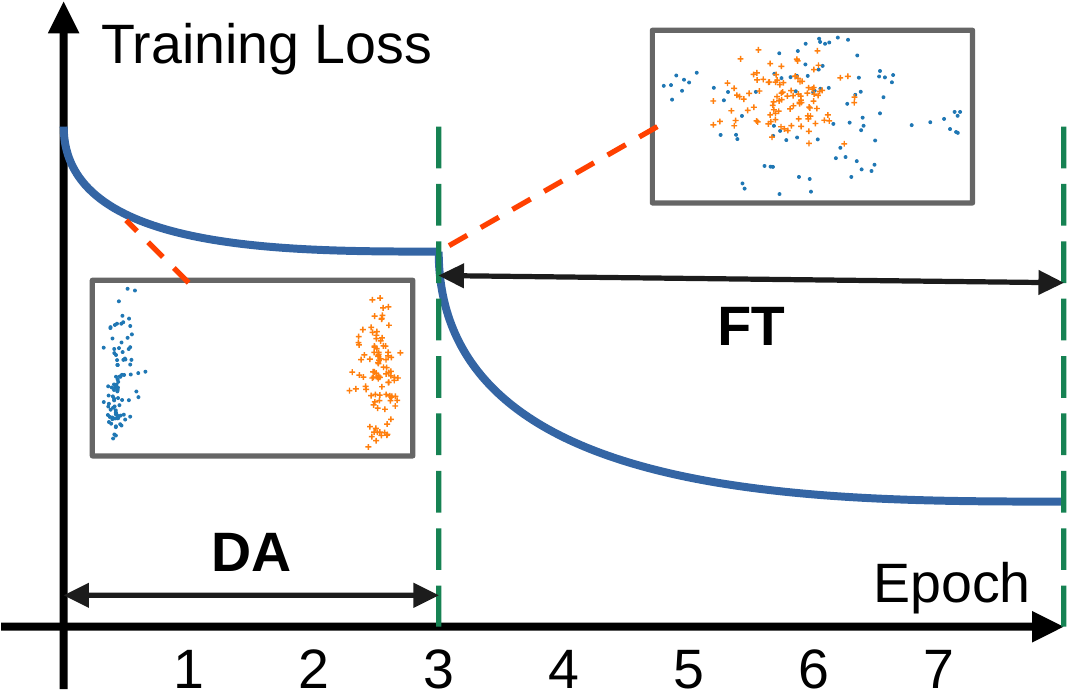}
      \caption{with alignment (\approach{})}
      \label{fig:intro.training_loss.daft}
    \end{subfigure}
    \caption{The representation collapse while training heterogeneous dual-encoders. The embeddings are visualized in two-dimensional PCA plots, where queries are orange and documents are blue. Without prior alignment (\cref{fig:intro.training_loss.no_daft}), the training loss briefly decreases before increasing and approaching a constant limit, which corresponds to the dimensional collapse. Using \approach{} (\cref{fig:intro.training_loss.daft}), the query encoder is aligned to the document encoder first before both encoders are fine-tuned in order to avoid the collapse.}
    \label{fig:intro.training_loss}
\end{figure}
A natural design decision to improve the efficiency of the dual encoders is to use a lightweight query encoder.
Documents are longer pieces of well-formed text that are known to be encoded well by transformer models.
Queries are often short and concise, suggesting that query encoders need not be overly complex.
There is evidence showing that it is beneficial to adapt the query and document encoders to their respective characteristics~\cite{jung2022semi}.
Extreme implementations of this are by \citet{zhuang2021tilde,zhuang2021fast}, who use sparse parameter-free query encoders with little drop in performance, but these models are only used for re-ranking.
We are not aware of any approaches using parametric yet lightweight neural query encoders.
In this paper, our aim is to train a lightweight query encoder to reduce the query encoding latency.

Reducing the complexity (in terms of architecture or number of parameters) of the query encoder without modifying the document encoder requires the model to have a \emph{heterogeneous} architecture. 
This means that, instead of having identical encoders, two separate and entirely different query and document encoder models have to be used. 
However, we show in this work that training heterogeneous dual-encoders using contrastive loss functions, that are commonly used for training homogeneous encoders, is in fact quite challenging.
Specifically, we find that a \emph{representation collapse} occurs, where both encoders slowly \emph{degenerate} and end up outputting the same constant trivial representation vector either on all (\emph{complete collapse}) or a subset (\emph{dimensional collapse}) of its dimensions for arbitrary inputs. 
\Cref{fig:intro.training_loss.no_daft} shows that standard contrastive training results in query and document vectors collapsing into a constant vector.
This causes the training loss to \emph{increase} after an initial decrease.
A similar problem has been observed in the field of visual representation learning and generative adversarial networks studied under \emph{mode collapse}~\cite{lala2018evaluation}.

Our first contribution in this paper is to show under what conditions the representation collapse occurs.
To this end, we show that the encoders, when they do not share initializations, are susceptible to collapsing due to a \emph{distribution mismatch} caused by their inherent language models.
Second, towards avoiding the representation collapse, we propose a novel yet simple two-stage training approach, \approach{}, which \emph{aligns} the distributions of the two encoders prior to fine-tuning in order to prevent them from collapsing. 
Aligning the query and document distributions based on their vector space geometry is key to ensuring that the query and document vectors are close enough to avoid collapsing (cf. \cref{fig:intro.training_loss.daft}).
To further improve the efficiency of the alignment stage, we propose fast estimations of the distribution dissimilarities to come up with a novel early stopping criterion.

Using \approach{}, we are able to train efficient heterogeneous dual-encoders that use \textit{lightweight query encoders}. 
Our query encoders achieve speed-ups of up to 4x with only small drops in performance.

In sum, we make the following contributions:
\begin{itemize}
    \item We establish the problem of \textit{collapsing representations} in heterogeneous neural retrieval models.
    \item We propose \approach{}, a simple two-stage training approach for alleviating the representation collapse problem and successfully training heterogeneous dual-encoder models.
    \item We conduct extensive experimental evaluation and show that \approach{} can be used to train efficient retrieval models using lightweight query encoders.
\end{itemize}
Our code will be open-sourced.

\section{Related Work}
\label{sec:related_work}
We divide the related work section into two parts -- \emph{neural retrieval and ranking} and the \emph{collapse of representations}.

\subsection{Neural Retrieval and Ranking}
\label{sec:related_work.retrieval_ranking}
Dense retrievers employ neural self-attention-based language models to compute vector representations of queries and documents. The original \dpr{} approach by \citet{karpukhin2020dense} paved the way for a number of improved dense retrieval models, with most improvements being made with respect to the training method. \ance{}~\cite{xiong2021approximate} is trained using hard negatives obtained from an ANN index that is updated asynchronously during the training. \tct{}~\cite{lin2021batch} employs knowledge distillation from the \colbert{} model~\cite{khattab2020colbert}. \citet{lindgren2021efficient} maintain a \emph{negative cache} to efficiently obtain the most impactful negative examples during training. \citet{zhan2021Optimizing} start by training the model with random negatives in order to improve its stability and subsequently use dynamic hard negatives for better ranking performance. \citet{gao2021condenser} argue that existing dense language models are not optimal for retrieval and propose the \condenser{} architecture, which, in combination with LM pre-training, allows for better aggregation of information in a single vector representation. This architecture is further improved with \cocondenser{}~\cite{gao2022unsupervised}.

The above-mentioned implementations base their encoders on \bert{}-base or similar transformer models, where the query and document encoder use shared weights, i.e.\, in a Siamese fashion. \citet{jung2022semi} use a \emph{semi-Siamese} setting, where the encoders do share parameters as well, but they are adapted to their specific role (query or document encoding) using \emph{light fine-tuning} methods. We are not aware of any approaches that employ \emph{heterogeneous models}, where the two encoders do not share the same model architecture and initial weights.

Hybrid retrievers combine sparse term-matching-based models~\cite{maron1960relevance} with dense neural models. Typically, two sets of documents are retrieved and subsequently combined. \clear{}~\cite{gao2021complement} takes the sparse component into account in the loss margin during the training phase of the dense model. \coil{}~\cite{gao2021coil} builds dense representations of terms using a deep language model in a pre-processing step.

Work on efficiency for dual-encoder-based models is limited to re-ranking so far. \tildeone{}~\cite{zhuang2021tilde} replaces the query encoder with a probabilistic model. \tildetwo{}~\cite{zhuang2021fast} uses exact contextualized term matching to reduce the memory requirements. \citet{leonhardt2022efficient} propose vector forward indexes that allow for efficient interpolation-based re-ranking using dual-encoders.

\subsection{Collapse of Representations}
\label{sec:related_work.collapse}
The problem of the dimensional or complete collapse of vector representations by encoder models has been well studied in the field of computer vision. Specifically, encoder models are employed to learn vector representations of visual inputs in an unsupervised fashion by training the model to output similar representations for multiple augmented versions of the same original image~\cite{chen2021exploring}. In order to ensure that the model does not output the same vector for every input (\emph{complete collapse} of the vector representations), contrastive loss functions~\cite{chen2020simple}, \emph{stop-gradient} approaches~\cite{chen2021exploring} and specific regularization terms~\cite{bardes2022vicreg} have been used. However, \citet{jing2022understanding} recently found that even contrastive loss functions are unable to entirely prevent the representations from collapsing to a subset of the dimensions in the vector space (\emph{dimensional collapse} of the vector representations). Overall, the problem of collapsing representations and the approaches to solve it are not entirely understood so far~\cite{zhang2022how}.

A similar issue exists in the field of generative adversarial networks (GANs), which have a common limitation known as \emph{mode collapse}, where the resulting generative model fails to capture some of the modes from the training data~\cite{lala2018evaluation}. This leads to reduced variability in the generated data. Several approaches have been proposed to alleviate this issue~\cite{srivastava2017veegan,bang2021mggan}.

In natural language processing, a related problem is known as \emph{anisotropy in language representations}~\cite{gao2021simcse}. Anisotropic embeddings are characterized as being concentrated within a small subspace (i.e.\ a \emph{cone}) of the embedding space, limiting their expressiveness.

To the best of our knowledge, the problem of collapsing representations has not been studied in the context of dual-encoder models for neural information retrieval as of yet.

\section{Collapse of Representations in IR}
\label{sec:collapse_ir}
In this section we briefly describe the architecture and training of dense retrieval models and introduce the problem of collapsing query and document representations caused by heterogeneous encoders.

\subsection{Dense Retrieval}
\label{sec:collapse_ir.dense_retrieval}
\emph{Dense retrievers} (DRs) are typically dual-encoder-based neural models, i.e.\ they employ a \emph{query encoder} $\zeta$ and a \emph{document encoder} $\eta$. The encoders map their corresponding inputs to fixed-size vector representations. Let a query $q = (t^q_1, ..., t^q_{ \vert q \vert })$ and a document $d = (t^d_1, ..., t^d_{ \vert d \vert })$ be sequences of tokens (e.g.\ subwords). The \emph{relevance score} of $d$ with respect to $q$ is then computed as
\begin{equation}
    \label{eq:collapse_ir.dense_retrieval.relevance_score}
    \Phi(q, d) = \zeta(q) \cdot \eta(d),
\end{equation}
where $\zeta(q), \eta(d) \in \mathbb{R}^a$. The dot product of the two vector representations encodes their similarity in the vector space. Note that $\zeta(q)$ and $\eta(d)$ are often normalized, such that $\Phi(q, d) \in [-1; 1]$.

If documents in the corpus are long, they are often split into passages prior to indexing. The relevance score for a document is then an aggregation of the corresponding passage scores. For example, the following approach is referred to as \emph{maxP}~\cite{dai2019deeper}:
\begin{equation}
    \label{eq:collapse_ir.dense_retrieval.maxp}
    \phi(q, d) = \max_{p_i \in d} (\zeta(q) \cdot \eta(p_i))
\end{equation}

In order to create a dense index $I_\mathcal{D}$ for a corpus $\mathcal{D}$, the vector representations of all documents (or passages) in the corpus are pre-computed and stored:
\begin{equation}
    I_\mathcal{D} = \{ \eta(d) \mid d \in \mathcal{D} \}
\end{equation}

\subsubsection{Query Processing}
\label{sec:collapse_ir.dense_retrieval.query_processing}
Retrieving the top-$k$ documents for an incoming query $q$ entails two steps: First, the query representation $\zeta(q)$ is computed. Afterwards, the $k$ highest scoring documents, i.e.\ the ones most similar to $q$, are returned as $K^q$. This is equivalent to a $k$-nearest-neighbor ($k$NN) search operation:
\begin{equation}
    K^q = \kargmax_{1 \leq i \leq \vert \mathcal{D} \vert } (\zeta(q) \cdot \eta(d_i))
\end{equation}
As dense indexes can become very large depending on size of the corpus and dimension $a$ of the representations, an exact $k$NN search is usually not feasible. Instead, index compression techniques such as product quantization are often used to conduct \emph{approximate} nearest neighbor (ANN) search with GPU acceleration~\cite{johnson2021billion}.

\subsubsection{Training}
\label{sec:collapse_ir.dense_retrieval.training}
The training objective of dense retrieval models usually involves a contrastive loss function. The idea is to simultaneously move relevant documents towards the query and push irrelevant documents away from it. Let the training set consist of query-document pairs $(q_i, d_i^+)$, where $d_i^+$ is a document that is relevant to $q_i$. The loss function can be expressed as follows:
\begin{equation}
    \label{eq:collapse_ir.dense_retrieval.training.loss}
    \mathcal{L}(q, d^+, D^-) =
        -\log
            \left(
                \frac
                    {\exp \left(\Phi(q, d^+; \theta) / \tau \right)}
                    {\sum_{d \in D^- \cup \{d^+\}} \exp \left(\Phi(q, d; \theta) / \tau \right)}
            \right)
\end{equation}
The hyperparameter $\tau$ denotes the \emph{temperature}, $D^-$ is a set of negative examples for $q$ and $\theta$ is the set of trainable parameters. Ideally, $D^-$ would contain all documents in the corpus, however, this would be computationally infeasible. Instead, a smaller set of negatives can be taken, for example, from within the same training batch~\cite{karpukhin2020dense}, an asynchronous index~\cite{xiong2021approximate} or a streaming cache~\cite{lindgren2021efficient}. The choice of negatives is important for the performance of the resulting model~\cite{zhan2021Optimizing}.

\subsection{Heterogeneous Dual-Encoders}
\label{sec:collapse_ir.heterogeneous_dual_encoders}
In practice, dual-encoder-based dense retrieval models, as described in \cref{sec:collapse_ir.dense_retrieval}, employ Siamese encoders, meaning that the query and document encoder are either the same model ($\zeta = \eta$) or share the same architecture and initialization. Examples are \ance{}~\cite{xiong2021approximate} and \tct{}~\cite{lin2021batch}. There are benefits to this approach; most prominently, the overall number of parameters is roughly cut in half, which has positive implications on the memory footprint during training. However, it also comes at a cost due to some undesirable properties:
\begin{enumerate}
    \item The encoders may be unable to fully adapt to the characteristics of their respective inputs. For example, queries are usually short and concise whereas documents are longer and more complex~\cite{jung2022semi}.
    \item The query encoder has the same number of parameters as the document encoder by design. This can be undesirable, as query representations can not be pre-computed. Thus, in order to keep the query processing latency at a minimum, the query encoder should be as lightweight as possible.
\end{enumerate}
The points above suggest that \emph{heterogeneous} dual-encoder models, where the query encoder is smaller (in terms of the number of parameters) than the document encoder, can improve the retrieval efficiency, while maintaining comparable overall effectiveness.

\begin{figure}
    \includegraphics[width=\linewidth]{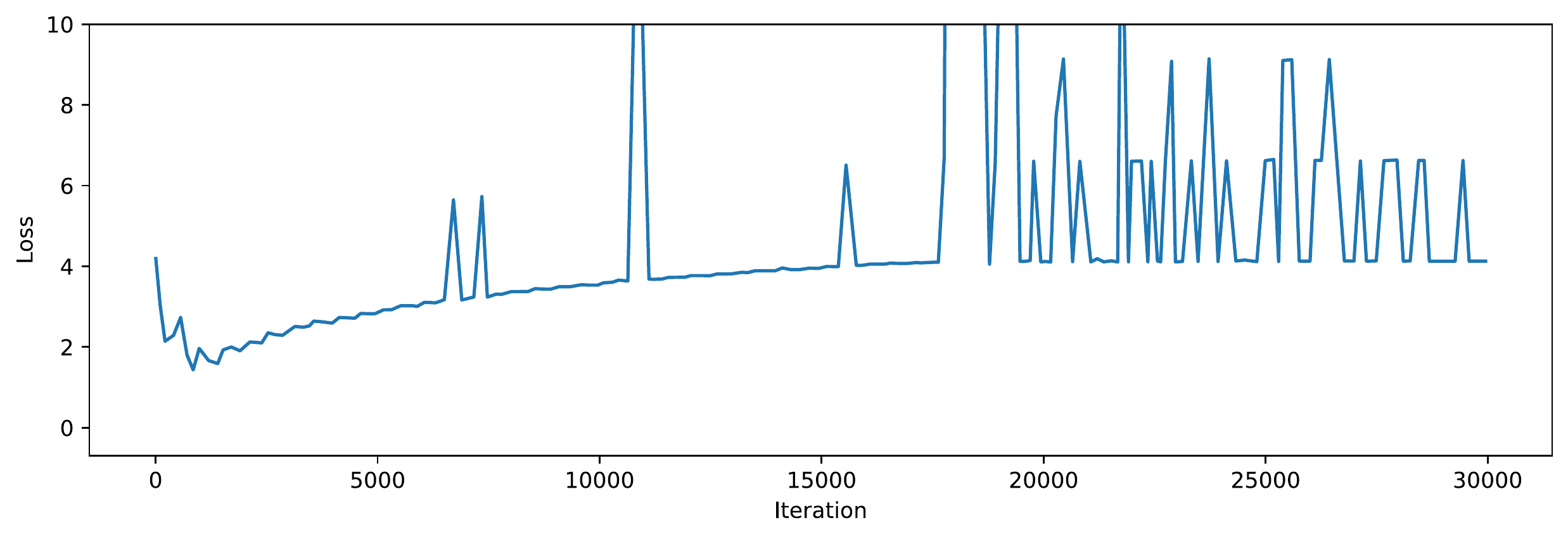}
    \caption{The training loss over time resulting from a heterogeneous model, where the query encoder uses a 2-layer \bert{} model and the document encoder uses a 12-layer \bert{} model.}
    \label{fig:collapse_ir.heterogeneous_dual_encoders.train_loss_collapse}
\end{figure}
\begin{figure}
    \begin{subfigure}{.49\linewidth}
      \includegraphics[width=\textwidth]{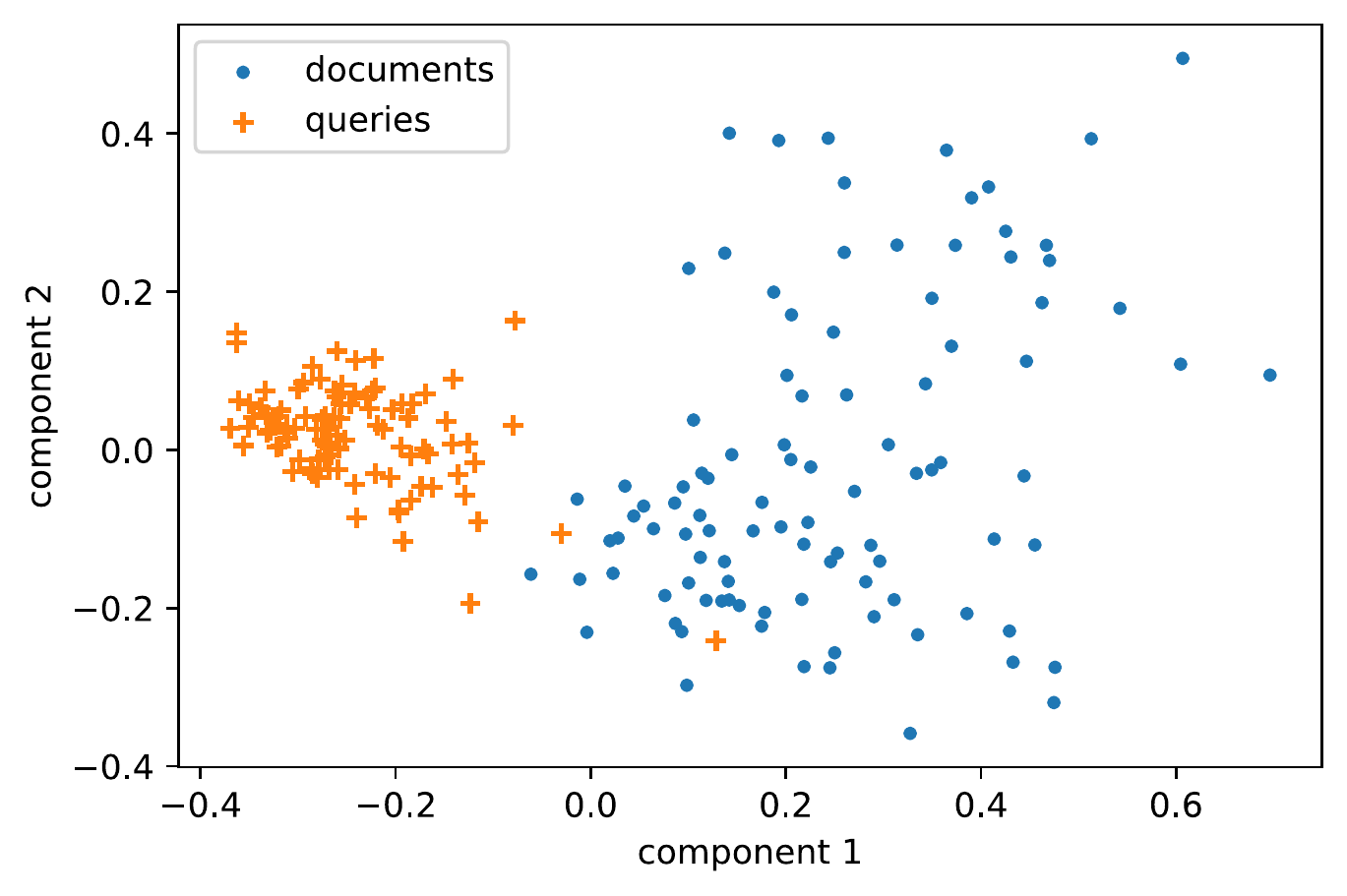}
      \caption{$\zeta = \eta$}
      \label{fig:collapse_ir.heterogeneous_dual_encoders.init_distr.homogeneous}
    \end{subfigure}
    \begin{subfigure}{.49\linewidth}
      \includegraphics[width=\textwidth]{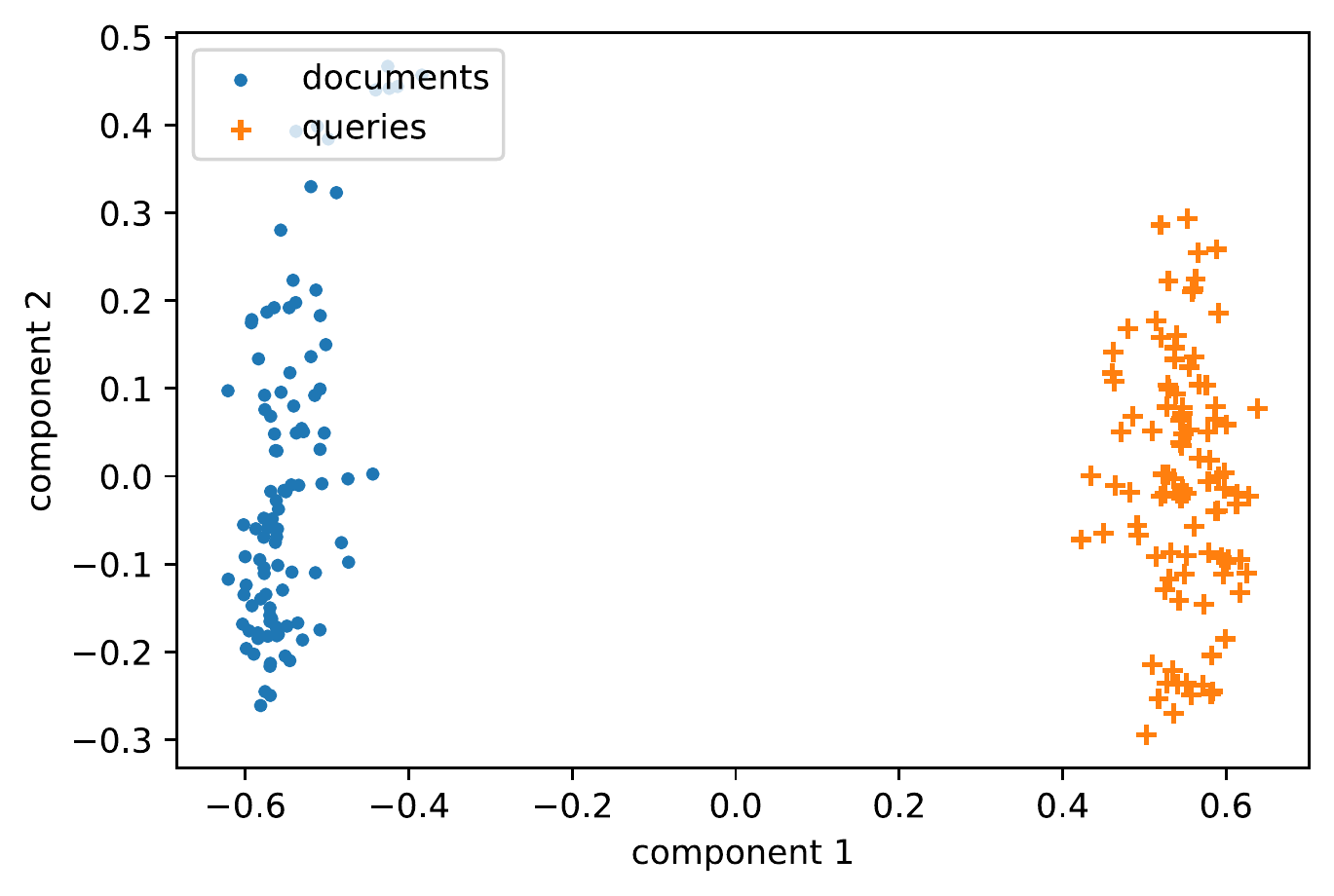}
      \caption{$\zeta \neq \eta$}
      \label{fig:collapse_ir.heterogeneous_dual_encoders.init_distr.heterogeneous}
    \end{subfigure}
    \caption{The initial distributions (i.e.\ before training) of (sampled) query and document representations from the MS MARCO dev set in the vector space after applying PCA. The heterogeneous setting (\cref{fig:collapse_ir.heterogeneous_dual_encoders.init_distr.heterogeneous}) exhibits a notable distribution mismatch.}
    \label{fig:collapse_ir.heterogeneous_dual_encoders.init_distr}
\end{figure}
The training of heterogeneous models, however, is not as straight-forward and poses some additional challenges. Concretely, the contrastive loss as shown in \cref{eq:collapse_ir.dense_retrieval.training.loss} causes the query and document representations to slowly \emph{collapse} during training. Similar phenomena have been observed in representation learning for computer vision as well as GANs (\emph{mode collapse}). \Cref{fig:collapse_ir.heterogeneous_dual_encoders.train_loss_collapse} shows a manifestation of this in the training loss of a heterogeneous model. The loss decreases for a short time before increasing again and approaching a constant value of $\mathcal{L} \approx 4.17$. This implies that the query and document representations have \emph{collapsed}, i.e.\ both encoders output a (nearly) constant vector $v$ irrespective of the input. As a result, we get 
\begin{equation}
    \Phi(q, d) = \zeta(q) \cdot \eta(d) \approx v \cdot v = 1 \quad \forall q, d
\end{equation}
and the loss for $\vert D^- \vert = 64$ (the number of negative samples used during training) becomes
\begin{equation}
    \mathcal{L}(q, d^+, D^-) \approx -\log \left(\frac{\exp (1 / \tau)}{65 \exp (1 / \tau)} \right) \approx 4.17.
\end{equation}
Retrieval models that have collapsed become completely unusable, as each query-document pair gets assigned (approximately) the same score, causing rankings to become effectively random. Evaluation of these models yields metric scores close to zero.

By analyzing the heterogeneous encoders we identify a \emph{distribution mismatch} as the likely cause of this. \Cref{fig:collapse_ir.heterogeneous_dual_encoders.init_distr} illustrates how sets of randomly sampled queries and documents are represented in the vector space before the training (and hence the collapse). It shows how, in the homogeneous case (\cref{fig:collapse_ir.heterogeneous_dual_encoders.init_distr.homogeneous}), the representations of queries and documents are close to each other and overlap slightly. In the heterogeneous case (\cref{fig:collapse_ir.heterogeneous_dual_encoders.init_distr.heterogeneous}), however, queries and documents are mapped to their own individual clusters, which are far apart from each other. This is likely due to the random initialization and slight differences in the language model pre-training of the encoders. We hypothesize that the contrastive ranking loss is unable to alleviate the distribution mismatch, causing the representations to collapse during training. In the next section, we introduce a simple training approach for heterogeneous dense retrieval models.

\section{Distribution-Aligned Fine-Tuning}
\label{sec:approach}
\begin{algorithm}[t]
    \DontPrintSemicolon
    \SetKwFunction{EarlyStop}{early\_stop}
    \SetKwFunction{BackProp}{backprop}
    \SetKwFunction{NegativeSample}{negative\_sample}
    \SetKw{Not}{not}
    \SetKw{In}{in}
    \KwIn{query encoder $\zeta$, document encoder $\eta$, training data $D_t$, validation data $D_v$, number of epochs $N_e$}
    
    $\theta_\zeta \gets \text{trainable parameters of } \zeta$\;
    $\theta_\eta \gets \text{trainable parameters of } \eta$\;
    
    \tcp{Stage 1: Distribution Alignment}
    \While{\Not $\EarlyStop(\zeta, \eta, D_v)$}{
        \ForEach{$(q, d^+)$ \In $D_t$}{
            $D^- \gets \NegativeSample(q, D_t)$\;
            $\ell \gets \mathcal{L}(q, d^+, D^-)$\;
            $\theta_\zeta \gets \BackProp(\ell, \theta_\zeta)$\;
        }
    }
    
    \tcp{Stage 2: Fine-Tuning}
    \For{$e \gets 1$ \KwTo $N_e$}{
        \ForEach{$(q, d^+)$ \In $D_t$}{
            $D^- \gets \NegativeSample(q, D_t)$\;
            $\ell \gets \mathcal{L}(q, d^+, D^-)$\;
            $\theta_\zeta \gets \BackProp(\ell, \theta_\zeta)$\;
            $\theta_\eta \gets \BackProp(\ell, \theta_\eta)$\;
        }
    }

    \caption{\approach{}}
    \label{alg:approach.training_approach}
\end{algorithm}

As outlined in \cref{sec:collapse_ir.heterogeneous_dual_encoders}, the distribution mismatch of the two heterogeneous encoders causes a collapse of the query and document representations during training using a contrastive loss. In this section we introduce our simple two-step training approach, \emph{distribution-aligned fine-tuning} (\approach{}), which allows for the training of heterogeneous dual-encoder-based retrieval models. The two stages of the algorithm may be described as follows:
\begin{enumerate}
    \item \textbf{\underline{D}istribution \underline{A}lignment}: In the first stage, the query encoder is aligned to the document encoder. Concretely. the document encoder remains frozen (i.e.\ its weights are not updated during backpropagation) and only the query encoder is trained. This stage employs an \emph{early stopping} technique to stop as soon as the two encoders are \emph{sufficiently aligned}. The early stopping criteria are explained in detail in \cref{sec:approach.early_stopping}.
    \item \textbf{\underline{F}ine-\underline{T}uning}: In the second stage, both encoders are fine-tuned, i.e.\ the document encoder weights are included in the backward pass. This matches the standard fine-tuning approach of neural retrieval models.
\end{enumerate}
\Cref{alg:approach.training_approach} illustrates the approach. The two stages are applied consecutively: The alignment stage updates the query encoder weights until both encoders are aligned; the fine-tuning stage trains both encoders for a fixed number of epochs. During both stages, the contrastive loss as in \cref{eq:collapse_ir.dense_retrieval.training.loss} is used.

We also tried a slightly different training approach, where, instead of a dedicated alignment phase, we trained both encoders in an \emph{alternating} fashion, i.e.\ only one encoder is updated for each training batch. We found that this approach also worked, but the overall training process was slower.

\subsection{Negative Sampling}
\label{sec:approach.negative_sampling}
As \approach{} employs a contrastive loss, a set of negative documents, $D^-$, has to be sampled for each query $q$ during training (cf.\ \cref{alg:approach.training_approach}). The sampling technique is independent of the algorithm itself, i.e.\ for simplicity, an in-batch strategy could be used.

For efficiency reasons, we adopt the \emph{negative cache} strategy proposed by \citet{lindgren2021efficient} for our experiments. The algorithm works by maintaining a \emph{streaming cache} of document representations during training, the size of which is controlled by a \emph{cache fraction} parameter $\alpha$. After each training iteration, a \emph{refresh fraction} $\rho$ of the cache is updated, i.e.\ the document representations are re-computed. In order to sample the set of negative documents $D^-$ for a query, a weighted Gumbel-max sample is taken from the cache, where the weights are computed as the query-document relevance scores (cf.\ \cref{eq:collapse_ir.dense_retrieval.relevance_score}). For further details, we refer the reader to the original paper.

\subsection{Early Stopping}
\label{sec:approach.early_stopping}
\begin{figure}
    \begin{subfigure}{.475\linewidth}
      \includegraphics[width=\textwidth]{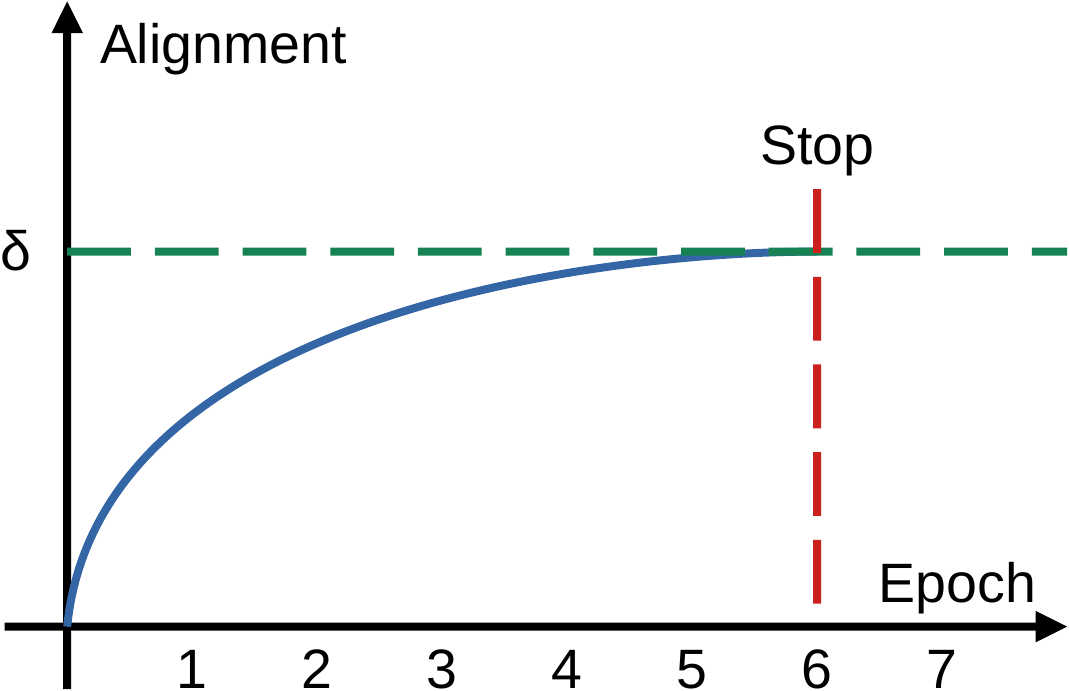}
      \caption{threshold $\delta$}
      \label{fig:approach.early_stopping.early_stopping.delta}
    \end{subfigure}
    \hfill
    \begin{subfigure}{.475\linewidth}
      \includegraphics[width=\textwidth]{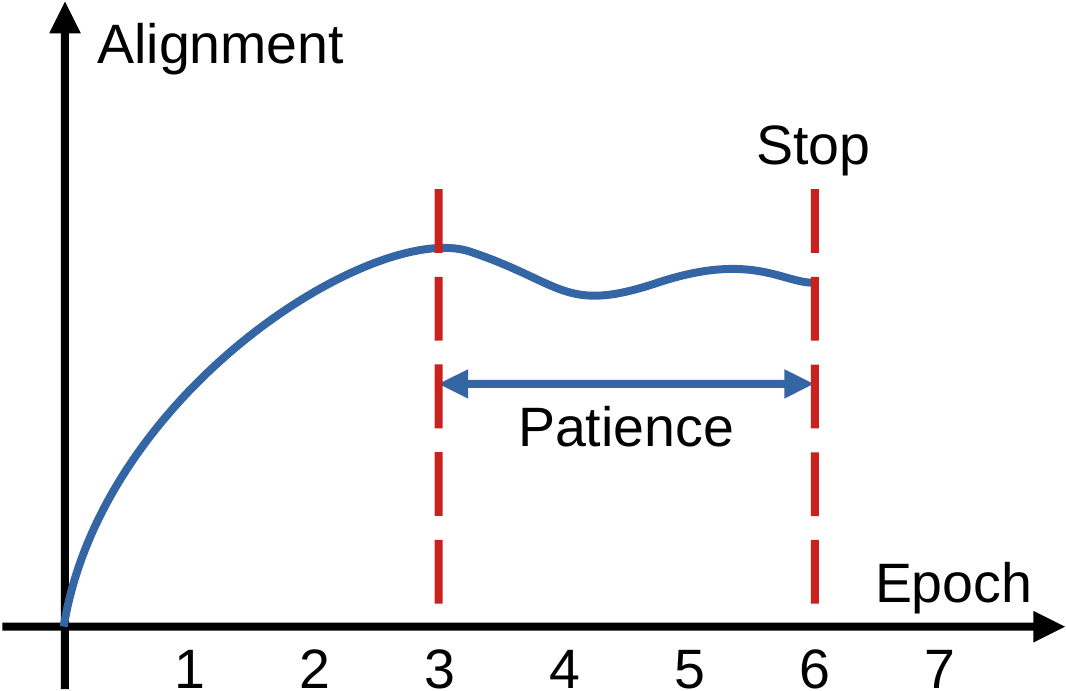}
      \caption{patience}
      \label{fig:approach.early_stopping.early_stopping.patience}
    \end{subfigure}
    \caption{Early stopping during the alignment stage of the \approach{} algorithm. As the alignment of the two encoders exceeds a threshold $\delta$ (\cref{fig:approach.early_stopping.early_stopping.delta}) or hasn't improved for a defined number of epochs (\cref{fig:approach.early_stopping.early_stopping.patience}), the process is stopped.}
    \label{fig:approach.early_stopping.early_stopping}
\end{figure}
The first stage of the \approach{} algorithm aligns the two encoders such that their distributions are sufficiently close to one another in the vector space in order to avoid a collapse of the representations during the second stage. Consequently, it is necessary to (1) come up with a way to measure the \emph{degree of alignment} after each epoch over the training data and (2) determine a suitable threshold $\delta$ to be exceeded before advancing to the next stage. Note that, usually, early stopping is based on the \emph{improvement} of a validation metric, i.e.\ the training is stopped once the validation score does not show improvements over a certain number of epochs (\emph{patience}). In the case of alignment, however, we noticed that the validation score (degree of alignment) generally slowly approaches a constant limit; thus, we employ the aforementioned threshold $\delta$ as an early stopping criterion. The process is illustrated in \cref{fig:approach.early_stopping.early_stopping}.

This corresponds to the early stopping condition in the first stage of \cref{alg:approach.training_approach}. In this section we introduce and discuss two ways of measuring alignment.

\subsubsection{ANN Retrieval}
\label{sec:approach.early_stopping.ann}
The most straightforward way of measuring the degree of alignment is by simply evaluating the model using ranking metrics. This approach is akin to evaluating the final model after the complete training process and involves indexing the collection (\cref{sec:collapse_ir.dense_retrieval}), performing ANN search using validation queries (\cref{sec:collapse_ir.dense_retrieval.query_processing}) and finally computing ranking metrics.

While the resulting performance in terms of ranking metrics is significantly lower than usual due to the fact that only the query encoder has been trained, this approach works well as an early stopping criterion, as the results are very accurate due to the model being evaluated directly. It is, however, extremely inefficient in terms of time and computational resources, as the whole collection needs to be re-indexed frequently. Thus, we propose an alternative approximate approach in the next section.

\subsubsection{Estimating Alignment}
\label{sec:approach.early_stopping.kl}
The Kullback-Leibler divergence~\cite{kullback1951on} (KL divergence) measures the \emph{dissimilarity} of two probability distributions $P$ and $Q$:
\begin{equation}
    \operatorname{KL}(P \Vert Q) = \int_{\mathbb{R}^a} p(x) \log \frac{p(x)}{q(x)}dx \geq 0
\end{equation}

We use an estimate of the KL divergence for vectorial data for the early stopping condition during the alignment stage. Let $\mathcal{X} = \{x_i\}_{i=1}^n$ and $\mathcal{X}' = \{x'_i\}_{i=1}^m$ be i.i.d.\ samples drawn from $p(x)$ and $q(x)$, respectively, with $x_i, x'_i \in \mathbb{R}^a$. The estimator proposed by \citet{perez2008kullback} is computed as
\begin{equation}
    \label{eq:approach.early_stopping.kl.vector_kl_estimate}
    \operatorname{KL}_k(P \Vert Q) = -\frac{a}{n} \sum_{i=1}^n \log \frac{r_k(x_i)}{s_k(x_i)} + \log \frac{m}{n - 1},
\end{equation}
where $r_k(x_i)$ and $s_k(x_i)$ are the Euclidian distances to the $k$-th nearest neighbor of $x_i$ in $\mathcal{X} \setminus x_i$ and $\mathcal{X}'$, respectively. $\operatorname{KL}_k(P \Vert Q)$ is proven to converge almost surely to $\operatorname{KL}(P \Vert Q)$.

Setting $k=1$, we use the estimate in \cref{eq:approach.early_stopping.kl.vector_kl_estimate} as our early stopping condition, where $\mathcal{X}$ corresponds to representations output by the document encoder and $\mathcal{X}'$ corresponds to representations output by the query encoder. The inputs of both encoders are queries sampled from the validation set. This allows us to approximate the degree of alignment of the query encoder without expensive re-indexing of the entire collection and performing ANN retrieval.

\section{Experimental Setup}
\label{sec:setup}
In this section we briefly describe the setup of our experimental evaluation. This includes details about datasets, model architecture, training and baselines.

\subsection{Datasets and Baselines}
\label{sec:setup.datasets_baselines}
Our models are evaluated on the \textit{TREC Deep Learning track} test set from the year 2019. 
It comprises 43 queries with corresponding relevance judgments over the MS MARCO v1 corpora. We evaluate on both the passage and document ranking task, however, our models are only trained on the passage corpus. For document ranking, we use the maxP approach (cf.~\cref{eq:collapse_ir.dense_retrieval.maxp}).

We compare our models mainly to other retrieval baselines; negative sampling methods include naive approaches, i.e.\ in-batch, random or BM25 negatives, as well as more involved techniques such as \ance{}~\cite{xiong2021approximate} and \staradore{}~\cite{zhan2021Optimizing} or pre-training methods like \condenser{}~\cite{gao2021condenser} and \cocondenser{}~\cite{gao2022unsupervised}.

\subsection{Model Architecture}
\label{sec:setup.architecture}
The specific model architecture we use for our dual-encoder setup is inspired by \citet{lindgren2021efficient}. Recall the computation of the query-document relevance score from \cref{eq:collapse_ir.dense_retrieval.relevance_score}. The two encoders $\zeta$ and $\eta$ employ Transformer-based \bert{} models, $\operatorname{LM}_\zeta$ and $\operatorname{LM}_\eta$, where the outputs corresponding to the \texttt{CLS} tokens are used to obtain initial vector representations of queries and documents, respectively. These representations are first fed through a \textit{shared linear projection layer} and then $L_2$-normalized:
\begin{align}
    \zeta(q) &= \left\Vert W \operatorname{LM}_\zeta(q) + b \right\Vert_2 \\
    \eta(d) &= \left\Vert W \operatorname{LM}_\eta(d) + b \right\Vert_2
\end{align}
$W \in \mathbb{R}^{a' \times a}$ and $b \in \mathbb{R}^a$ are trainable parameters corresponding to the shared projection. We empirically find that the shared projection is necessary to prevent the model from collapsing.

\subsection{Training and Evaluation Details}
\label{sec:setup.details}
Our models and training procedures are implemented using the \texttt{PyTorch}\footnote{\url{https://pytorch.org/}} framework and the \texttt{transformers}\footnote{\url{https://huggingface.co/transformers}} library. Pre-trained language models, such as \bert{}, are obtained from the \emph{Hugging Face Hub}\footnote{\url{https://huggingface.co/models}} platform. We use two NVIDIA Tesla V100 (32G) GPUs with 16-bit floating point precision for training. Latency measurements are taken solely on CPU using an Intel Xeon Silver 4210. Tokenizer latency is ignored.

\subsubsection{Validation}
\label{sec:setup.details.validation}
We perform validation using the MS MARCO development sets. Our models are trained on the passage corpus and subsequently evaluated on both the passage and document ranking tasks. Validation is performed on the development set corresponding to the task. The document corpus is not used for training.

Validation differs for the two steps of the \approach{} algorithm. During the alignment stage, we employ the estimated KL divergence as described in \cref{sec:approach.early_stopping.kl} on the validation set in order to do early stopping: The alignment is stopped as soon as at least one of two conditions is met: The estimated KL divergence (1) falls below the threshold $\delta = 250$ or (2) fails to decrease for three subsequent epochs (patience). These conditions are illustrated in \cref{fig:approach.early_stopping.early_stopping}. Note that the second condition acts as a fail-safe to avoid infinite loops in case the threshold is never reached, however, in our experiments, this never happened. We empirically determine $\delta = 250$ using a single configuration of heterogeneous encoders and find that it generalizes well to all other models we used on the same corpus.

During the fine-tuning stage, we do not do early stopping. Instead, we fine-tune for $N_e = 10$ epochs and select the best checkpoint based on the validation performance using ANN retrieval (cf.\ \cref{sec:approach.early_stopping.ann}), i.e.\ we retrieve passages from the corpus for each query in the development set and compute the nDCG@10 value to select the best checkpoint to use for evaluation.

\subsubsection{Hyperparameters}
\label{sec:setup.details.hyperparameters}
For all of our models, we use an uncased \bert{}-base model with $L = 12$ encoder layers, $A = 12$ self-attention heads and $768$-dimensional hidden representations within the document encoder $\eta$ (cf.\ \cref{sec:setup.architecture}). The query encoders $\zeta$ use uncased \bert{} models with varying numbers of layers~\cite{turc2019well}. Query and document representations $\zeta(q)$ and $\eta(d)$ have $a = 512$ dimensions. The models are optimized using \textsc{AdamW}~\cite{loshchilov2018decoupled}, a learning rate of \num{1e-5}, 1000 warm-up steps and a batch size of 48 (24 per GPU).

The negative cache (cf.\ \cref{sec:approach.negative_sampling}) holds 512k documents across all GPUs. For each training instance, we sample $\vert D^- \vert = 64$ negative documents from the cache (cf.\ \cref{eq:collapse_ir.dense_retrieval.training.loss}). The temperature is set to $\tau = 1$. Following the original implementation by \citet{lindgren2021efficient}, we multiply all scores by a factor of $20$ during training.

\subsubsection{Metrics}
\label{sec:setup.details.metrics}
We mainly focus on efficiency, i.e.\ the query encoding latency (in milliseconds). We further report standard retrieval and ranking metrics such as recall, MRR and nDCG.

\section{Results}
\label{sec:results}
In this section we experimentally evaluate \approach{} by investigating the following research questions:
\begin{itemize}
    \item[\bf RQ1] Which encoder models are affected by the collapse of query and document representations and how effective is \approach{} in preventing this?
    \item[\bf RQ2] Can we use the \approach{} algorithm to train \emph{more efficient heterogeneous retrieval models} by reducing the size of the query encoder without compromising effectiveness?
\end{itemize}
Our focus lies mainly on \emph{preventing the collapse of representations} and thus \emph{training efficient heterogeneous dual-encoders}, i.e.\ relative performance instead of beating the state-of-the-art.

\subsection{The Effect of Encoder Models}
\label{sec:results.encoder_models}
\begin{table}
    \begin{tabular}{lccc} 
        \toprule
        &
        \multicolumn{2}{c}{With \approach{}} &
        \multirowcell{2.3}{Without\\\approach{}} \\
        \cmidrule(lr){2-3}
        Document encoder $\eta$                     & nDCG@10   & R@1k \\
        \midrule
        \bert{}-base \cite{devlin2019bert}          & 0.554     & 0.575	& \collapse{} \\
        \roberta{}-base \cite{liu2019roberta}       & 0.515     & 0.579	& \collapse{} \\
        \condenser{} \cite{gao2021condenser}        & 0.579     & 0.602	& \collapse{} \\
        \cocondenser{} \cite{gao2022unsupervised}   & 0.575     & 0.629	& \collapse{} \\
        \mpnet{}-base \cite{song2020mpnet}          & 0.461     & 0.547	& \collapse{} \\
        \bottomrule
    \end{tabular}
    \caption{Retrieval results using various document encoder models $\eta$ on \trecdlpsg{} trained with and without alignment (\approach{} algorithm). $\vcenter{\hbox{\collapse{}}}$ indicates a collapse of representations during training. The query encoder $\zeta$ is a 2-layer \bert{} model.}
    \label{tab:results.encoder_models.collapse}
\end{table}
In this experiment we establish and further emphasize the existence of the problem of collapsing representations during the training of dense retrieval models. Specifically, we investigate a wide range of language models to determine whether they are affected. The models mainly differ in terms of architecture, and pre-training technique and data. We aim to find out whether only specific architectures or pre-training techniques make encoder models susceptible to the collapse. Additionally, we investigate whether our proposed \approach{} approach is able to prevent the corresponding models from collapsing during the training. This corresponds to \textbf{RQ1}.

The results are illustrated in \cref{tab:results.encoder_models.collapse}. It presents the ranking performance of various document encoder models after being trained in combination with a 2-layer \bert{} query encoder (heterogeneous setting). Each combination is trained and evaluated with and without the distribution alignment step of the \approach{} algorithm (cf.\ \cref{sec:approach}). Most importantly, the results show that all of the language models we considered eventually collapsed during the training. Further, the alignment step of \approach{} is able to successfully prevent the collapse in each of the cases.

\begin{figure*}
    \begin{subfigure}{.33\linewidth}
      \includegraphics[width=\textwidth]{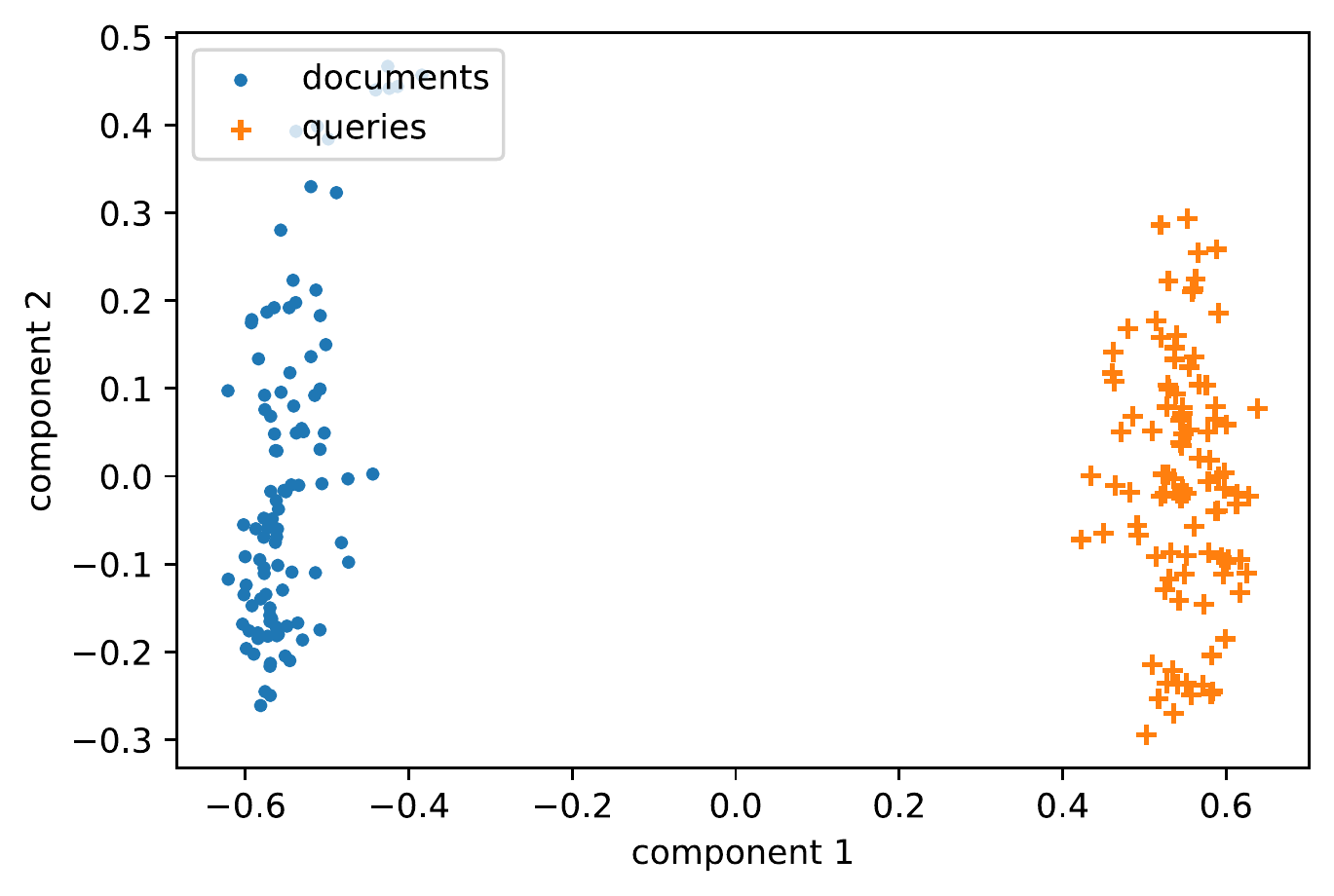}
      \caption{before fine-tuning}
      \label{fig:results.encoder_models.distributions.init}
    \end{subfigure}
    \begin{subfigure}{.33\linewidth}
      \includegraphics[width=\textwidth]{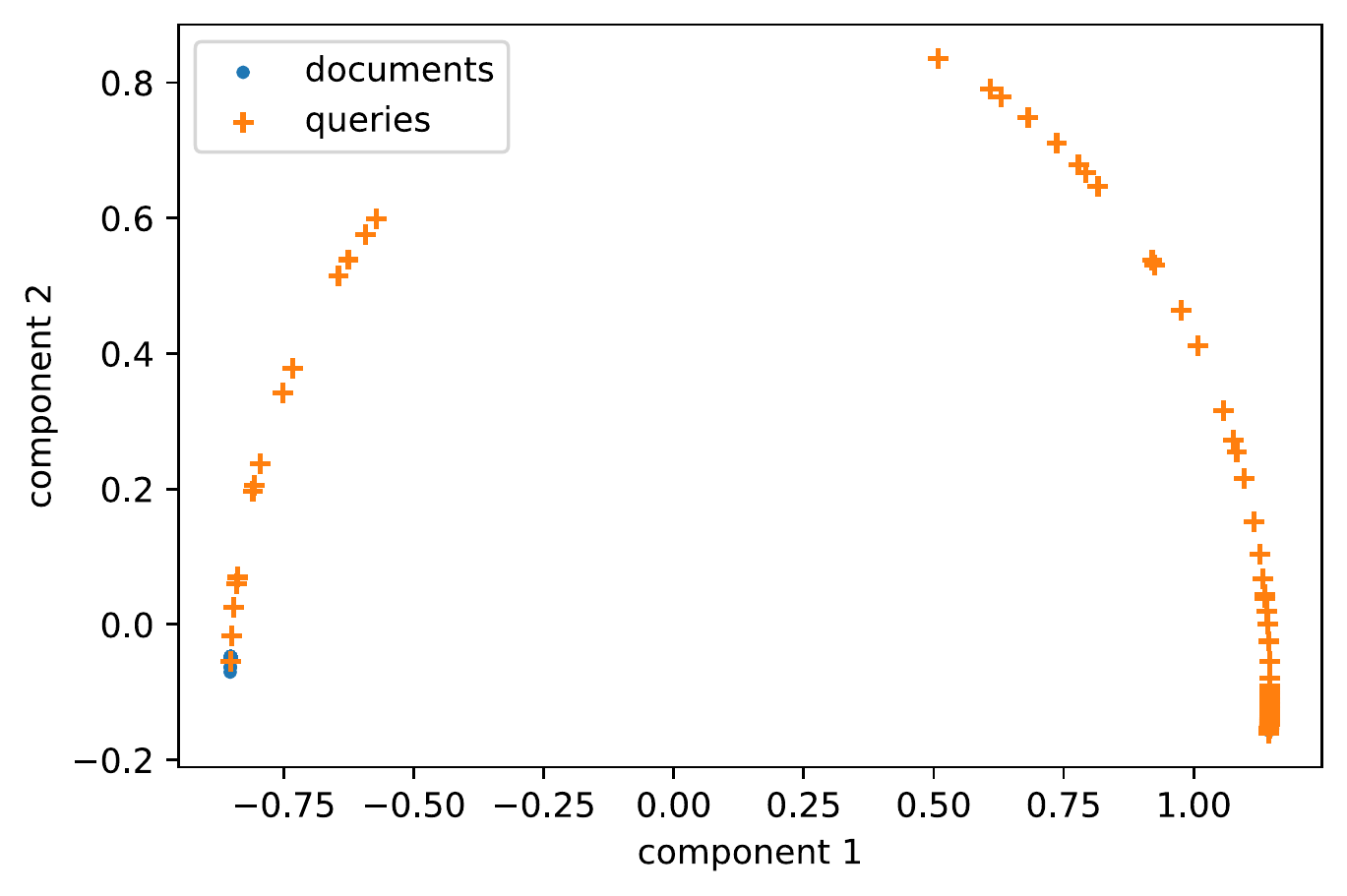}
      \caption{after fine-tuning without alignment}
      \label{fig:results.encoder_models.distributions.without_alignment}
    \end{subfigure}
    \begin{subfigure}{.33\linewidth}
      \includegraphics[width=\textwidth]{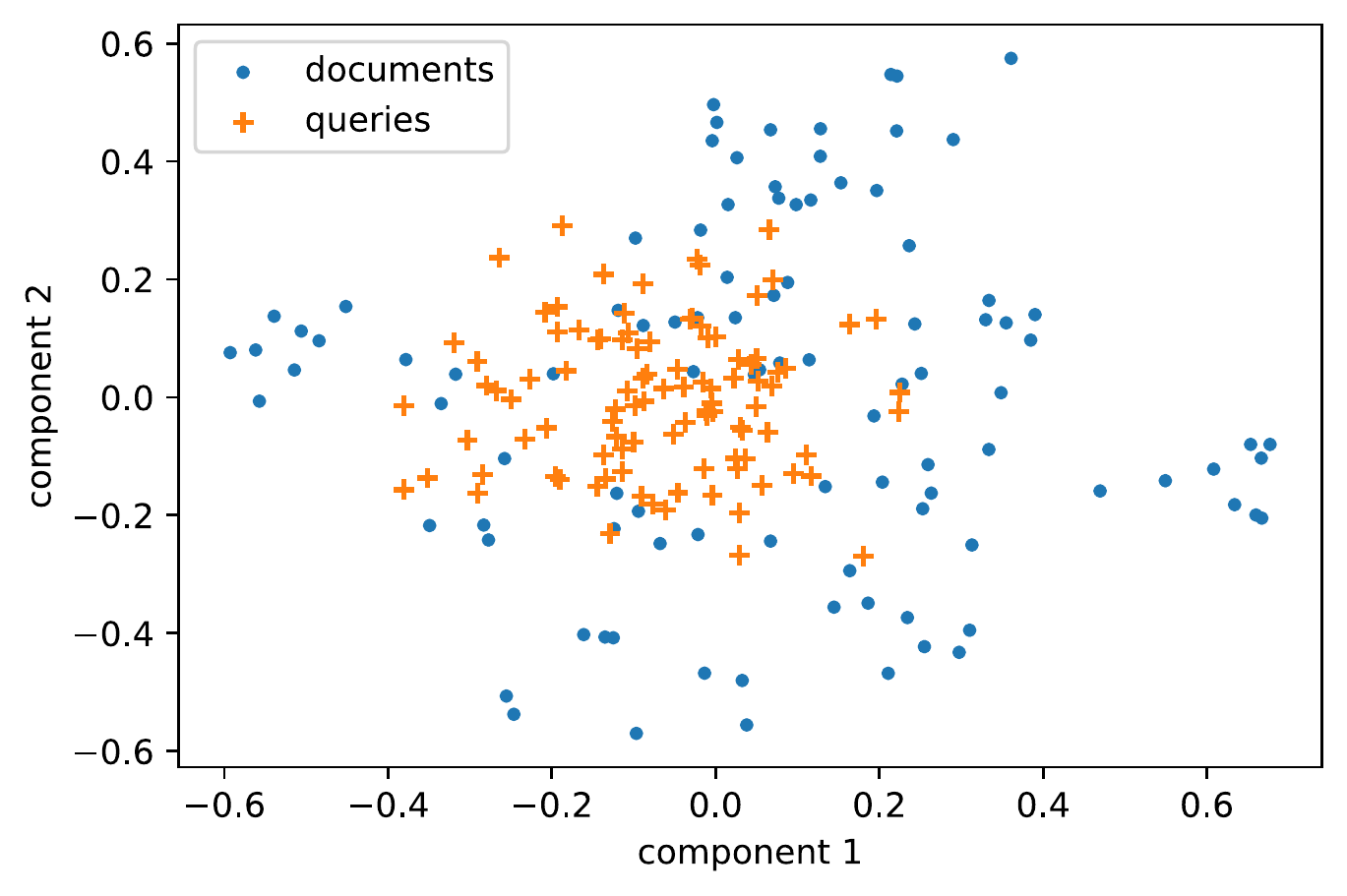}
      \caption{after alignment (\approach{})}
      \label{fig:results.encoder_models.distributions.with_alignment}
    \end{subfigure}
    \caption{The effect of fine-tuning a heterogeneous dual-encoder model with and without distribution alignment (PCA plots). The query encoder $\zeta$ is a 2-layer \bert{} model and the document encoder $\eta$ is a 12-layer \bert{} model (\bert{}-base). Initially (before any fine-tuning), the query and document representations are separated in two clusters (\cref{fig:results.encoder_models.distributions.init}). Fine-tuning directly without any prior alignment causes the encoders to collapse (\cref{fig:results.encoder_models.distributions.without_alignment}). Finally, the aligned encoders yield overlapping query and document representations (\cref{fig:results.encoder_models.distributions.with_alignment}) and can be fine-tuned without collapsing.}
    \label{fig:results.encoder_models.distributions}
\end{figure*}
In order to better understand this behavior, we plot a sample of the query and documents representations yielded by heterogeneous encoders in \cref{fig:results.encoder_models.distributions}. As before, due to the heterogeneous nature, the initial representations are grouped in two clusters with one corresponding to each encoder, respectively. The plots show how fine-tuning without a prior alignment step causes a dimensional collapse of the query encoder and a complete collapse of the document encoder. The alignment step of \approach{}, on the other hand, causes both encoders to output overlapping distributions, which can then be fine-tuned in the subsequent step.

\subsection{Efficient Query Encoders}
\label{sec:results.efficient_query_encoders}
\begin{table*}
        \begin{tabular}{llllllllll}
            \toprule
            &
            \multicolumn{2}{c}{\msmpsgdev{}} &
            \multicolumn{2}{c}{\trecdlpsg{}} &
            \multicolumn{2}{c}{\msmdocdev{}} &
            \multicolumn{2}{c}{\trecdldoc{}} &
            Latency \\
            \cmidrule(lr){2-3}
            \cmidrule(lr){4-5}
            \cmidrule(lr){6-7}
            \cmidrule(lr){8-9}
            \cmidrule(lr){10-10}
                                                            & MRR@10    & R@1k  & nDCG@10   & R@1k              & MRR@100   & R@1k  & nDCG@10   & R@1k              & ms \\ 
            \midrule
            \multicolumn{10}{l}{\bf \textsc{Negative Sampling}} \\
            In-Batch \cite{zhan2021Optimizing}              & 0.264     & -     & 0.583     & -                 & 0.320     & -     & 0.544     & -                 & 78.8 \\
            Random \cite{zhan2021Optimizing}                & 0.301     & -     & 0.612     & -                 & 0.330     & -     & 0.572     & -                 & 78.8 \\
            BM25 \cite{zhan2021Optimizing}                  & 0.309     & -     & 0.607     & -                 & 0.316     & -     & 0.539     & -                 & 78.8 \\
            \ance{} \cite{xiong2021approximate}             & 0.330     & 0.959 & 0.648     & -                 & -         & -     & 0.628     & -                 & 78.8 \\
            Negative Cache \cite{lindgren2021efficient}     & 0.322     & -     & 0.649     & -                 & -         & -     & -         & -                 & 79.2 \\
            \staradore{} \cite{zhan2021Optimizing}          & 0.347     & -     & 0.683     & -                 & 0.405     & -     & 0.628     & -                 & 78.8 \\
            \midrule
            \multicolumn{10}{l}{\bf \textsc{Distillation}} \\
            \tct{} \cite{lin2021batch}                      & 0.335     & 0.964 & 0.670     & 0.720             & -         & -     & -         & -                 & 79.1 \\
            \midrule
            \multicolumn{10}{l}{\bf \textsc{Pre-Training}} \\
            \condenser{} \cite{gao2021condenser}            & 0.366     & 0.974 & 0.698     & -                 & -         & -     & -         & -                 & 79.1 \\
            \cocondenser{} \cite{gao2022unsupervised}       & 0.382     & 0.984 & -         & -                 & -         & -     & -         & -                 & 79.1 \\
            \midrule
            \multicolumn{10}{l}{\bf \approach{} (ours)} \\
            {\tiny \texttt{[a]}} $\bert{}_{L=2}$            & 0.258     & 0.898 & 0.560     & 0.580             & 0.292     & 0.917 & 0.544     & 0.465             & 15.6 \impr{80.3} \\
            {\tiny \texttt{[b]}} $\bert{}_{L=4}$            & 0.266     & 0.919 & 0.580     & 0.607             & 0.296     & 0.917 & 0.563     & 0.453             & 29.4 \impr{62.8} \\
            {\tiny \texttt{[c]}} $\bert{}_{L=6}$            & 0.266     & 0.917 & 0.576     & 0.621\sig{a}      & 0.304     & 0.920 & 0.548     & 0.470             & 42.8 \impr{45.9} \\
            {\tiny \texttt{[d]}} $\bert{}_{L=8}$            & 0.270     & 0.924 & 0.557     & 0.599             & 0.298     & 0.921 & 0.582     & 0.482             & 55.1 \impr{30.3} \\
            {\tiny \texttt{[e]}} $\bert{}_{L=10}$           & 0.270     & 0.924 & 0.598     & 0.626\sig{a}      & 0.307     & 0.923 & 0.583     & 0.456             & 67.4 \impr{14.8} \\
            {\tiny \texttt{[f]}} $\bert{}_{L=12}$           & 0.284     & 0.940 & 0.592     & 0.677\sig{abcde}  & 0.314     & 0.941 & 0.595     & 0.532\sig{abcde}  & 79.1 \\
            \bottomrule
        \end{tabular}
        \caption{Passage and document retrieval performance of \approach{} models and various baselines. \approach{} models are heterogeneous, where the document encoder $\eta$ uses a 12-layer \bert{} model and the query encoder $\zeta$ uses a \bert{} model with fewer layers, as indicated by $L$ in the name. All baselines employ homogeneous models, the results are taking from the corresponding papers. Latency is reported as the time required to encode a query (on CPU) in milliseconds. Superscripts indicate statistically significant improvements ($p \leq 0.01$)~\cite{bassani2022ranx} on \trecdlpsg{} and \trecdldoc{}.}
        \label{tab:results.efficient_query_encoders.performance}
\end{table*}
\begin{figure}
    \includegraphics[width=\linewidth]{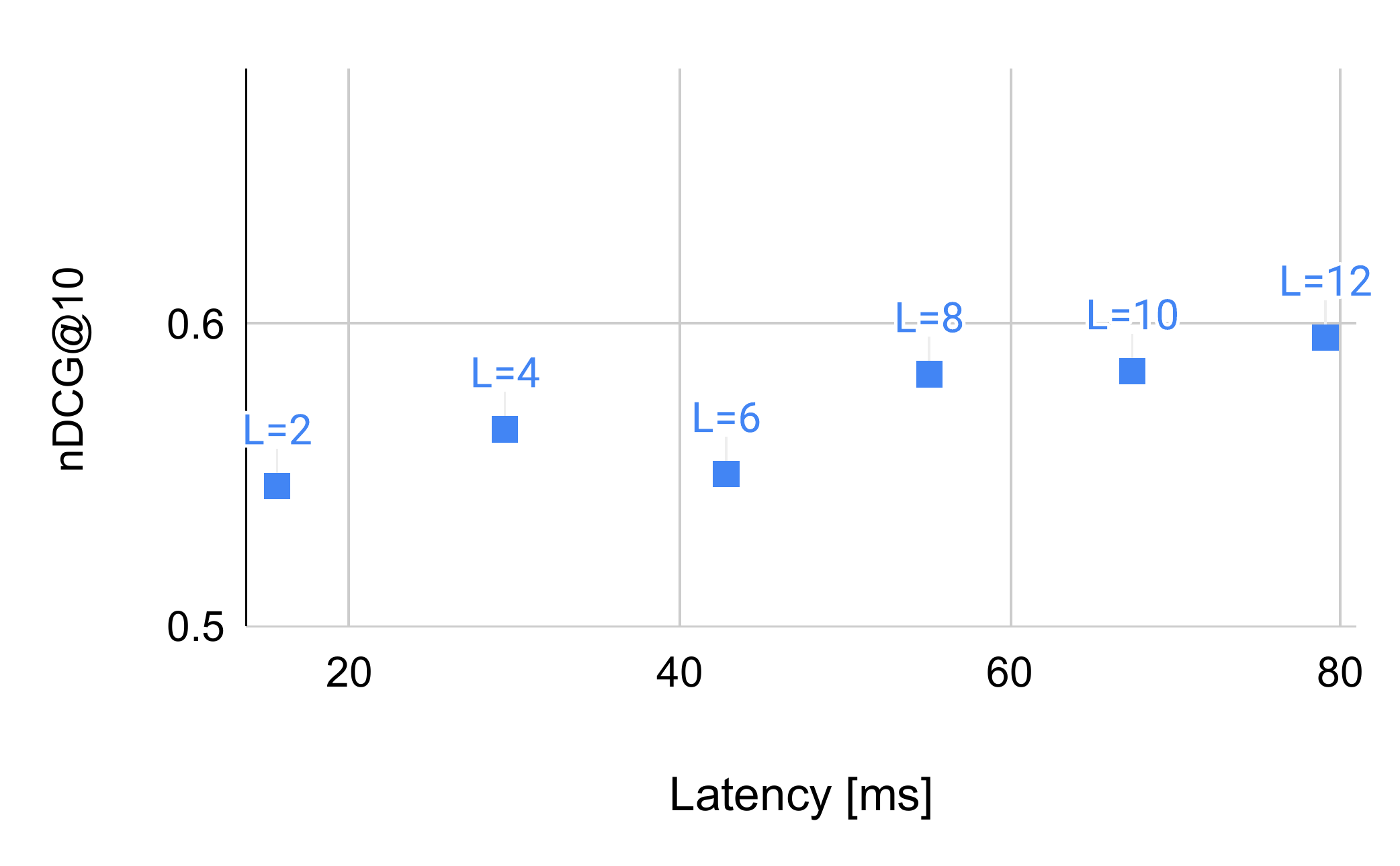}
    \caption{Retrieval performance with respect to query encoding latency of \approach{} models with varying size (complexity) on \trecdldoc{}.}
    \label{fig:approach.results.efficient_query_encoders.latency}
\end{figure}
In \cref{sec:results.encoder_models} we showed that the \approach{} algorithm can be used to prevent dimensional or complete collapse of representations during the training of heterogeneous retrieval models. In this experiment we train a number of heterogeneous models using \emph{efficient query encoders}, i.e.\ \bert{} models with successively fewer encoder layers, to answer \textbf{RQ2}.

\Cref{tab:results.efficient_query_encoders.performance} shows the results for passage and document retrieval of efficient \approach{}-based models along with a number of neural baselines. It is important to note here that \approach{} is in principle complementary to any of the baselines and could be combined with them. As a result, the performance of our models is heavily influenced by the fine-tuning (i.e.\ negative sampling) technique. Specifically, $\bert{}_{L=12}$ is a homogeneous model with the same encoders as the baselines, where the alignment step is not necessary; thus, the difference in performance is caused solely by the fine-tuning. For this reason, the \approach{} models should primarily be compared among each other rather than to the baselines. The fact that our models are unable to match the performance of the negative cache baseline (as reported by the original authors) could be due to subtle differences in the training data, validation or our implementation of the method.

We further plot the retrieval performance in terms of nDCG with respect to query encoding latency in \cref{fig:approach.results.efficient_query_encoders.latency}. The results show that, as expected, there is a trade-off between effectiveness (retrieval performance) and efficiency (query encoding latency), where the latency is directly correlated with the number of query encoder layers. The smallest model, $\bert{}_{L=2}$, achieves a speed-up of over 4x compared to $\bert{}_{L=12}$ (and thus the baseline models that have the same architecture), but the performance drops to some extent. However, we performed a two-sided paired Student's t-test ($p \leq 0.01$) and found that the drop in nDCG@10 is not statistically significant on both test sets for any number of encoder layers.

\section{Discussion and Conclusion}
\label{sec:discussion}
In this section we present and discuss additional insights we gained with respect to the training of heterogeneous models using the \approach{} algorithm. We further conclude and present an outlook on future work.

\subsection{Essential Model Components}
\label{sec:discussion.essential_components}
\begin{table}
    \begin{tabular}{cccc} 
        \toprule
        Baseline &
        \multicolumn{3}{c}{Omitted component} \\
        \cmidrule(lr){1-1} \cmidrule(lr){2-4}
        nDCG@10 & Shared proj.  & LM pre-training   & Self-att. \\
        \midrule
        0.560   & \collapse{}   & \collapse{}       & \collapse{} \\
        \bottomrule
    \end{tabular}
    \caption{Retrieval results on \trecdlpsg{} when query encoder components are omitted. The baseline corresponds to a 2-layer \bert{} query encoder and a 12-layer \bert{}-base document encoder. $\vcenter{\hbox{\collapse{}}}$ indicates a collapse of representations during training.}
    \label{tab:results.essential_components.overview}
\end{table}
We perform additional ablation studies in order to find out which parts (or properties) of the query encoder are essential to train heterogeneous dual-encoders without causing a dimensional or complete collapse of the representations. Concretely, the components in question are
\begin{enumerate}
    \item the shared linear projection layer (cf.\ \cref{sec:setup.architecture}), i.e.\ the query and document encoders simply consist of a language model and a normalization layer,
    \item the language model pre-training, i.e.\ the encoder weights are initialized randomly, and
    \item the self-attention mechanism, i.e.\ the query encoder becomes a simple average pooling operation over non-contextual token embeddings.
\end{enumerate}
\cref{tab:results.essential_components.overview} gives an overview over the omitted components and the corresponding performance (if applicable). It shows that, in fact, all components mentioned above are critical to the success of training a heterogeneous model in our setup with \approach{}, in other words, omitting any of them causes the encoders to collapse despite the alignment step. From these results, we draw the following (hypothetical) conclusions:
\begin{enumerate}
    \item At least a small shared component is required to \emph{kick-start} the alignment process properly.
    \item A certain \emph{quality} of the query representations is required for the encoders to work. Concretely, alignment and fine-tuning is not enough to compensate for the lack of language understanding (i.e.\ pre-training) or drastically simpler architectures (without self-attention).
\end{enumerate}

\subsection{Conclusion and Outlook}
\label{sec:discussion.conclusion}
In this paper we introduced the problem of collapsing representations in the information retrieval setting. Specifically, we showed that heterogeneous dual-encoder-based retrievers are prone to collapse when they are fine-tuned using a contrastive loss function.

We proposed \approach{}, a training approach for heterogeneous dual-encoders that prevents the collapse using an alignment stage prior to fine-tuning. We further utilized \approach{} to train more efficient neural retrievers by reducing the complexity of the query encoder.

Our initial findings suggest a number of ideas for possible future work; we have shown that query encoders require self-attention in order to work properly in our setup. This should be investigated in more detail with the goal of making the query encoder fully linear. On the other hand, the document encoder could be made larger to account for this loss in capacity. Finally, the impact of negative sampling on the problem of collapsing representations could be investigated.


\bibliographystyle{ACM-Reference-Format}
\bibliography{references}


\end{document}